\documentstyle[prl,aps,psfig]{revtex}
\begin{document}
\draft
%%%%%%%%%%%%%%%%%%
\twocolumn[\hsize\textwidth\columnwidth\hsize\csname
@twocolumnfalse\endcsname
%%%%%%%%%%%%%%%%%%

\widetext
\title{Nagaoka ferromagnetism in the two-dimensional infinite-$U$ Hubbard model}
\author{Federico Becca and Sandro Sorella}
\address{
Istituto Nazionale per la Fisica della Materia and
International School for Advanced Studies, Via Beirut 4,
34013 Trieste, Italy \\
}

\date{\today}
\maketitle
\begin{abstract}
We present different numerical calculations based on variational quantum
Monte Carlo simulations supporting a ferromagnetic ground-state for finite and small
hole densities in the two-dimensional infinite-$U$ Hubbard model. 
Moreover, by studying the  energies of different total spin sectors, these calculations 
strongly suggest that the paramagnetic phase is unstable
against a phase with a partial polarization for large hole densities $\delta \sim 0.40$ with 
evidence for a second-order transition to the paramagnetic large doping phase.
\end{abstract}
\pacs{75.10.Lp, 75.40.Mg}
%%%%%%%%%%%%%%%%%%
]
%%%%%%%%%%%%%%%%%%

\narrowtext
The Hubbard model was originally introduced to explain ferromagnetism 
in strongly correlated transition metals \cite{hubbard}:
\begin{equation}\label{hamilt}
H = -t \sum_{\langle i,j \rangle, \sigma} c^{\dag}_{i,\sigma} c_{j,\sigma}
+ U \sum_{i} n_{i,\uparrow} n_{i,\downarrow},
\end{equation}
where $\langle \; \rangle$ stands for nearest neighbors, $c_{i,\sigma}$
($c^{\dag}_{i,\sigma}$) destroys (creates) an electron with spin $\sigma$ at
site $i$, and $n_{i,\sigma}=c^{\dag}_{i,\sigma}c_{i,\sigma}$.
In the following all energies are measured in units of $t$.
In this simple and appealing Hamiltonian the nontrivial ingredient 
is that pairs of electrons with opposite spins that occupy the same site 
pay an on-site energy $U$.
For a large Coulomb interaction $U$, electrons can prefer to occupy only 
a single spin sector, up or down, thereby minimizing the strong Coulomb 
repulsion. Indeed, as Nagaoka showed in a milestone paper \cite{nagaoka},
a single hole in the infinite-$U$ Hubbard model on any finite bipartite 
cluster with periodic boundary conditions, for any dimension $d \ge 2$ has 
a fully polarized ferromagnetic ground-state, i.e., the ground-state has 
maximum spin $S$. Unfortunately, the Nagaoka theorem refers to a single point in the 
phase diagram, which is not thermodynamically significant. 
Therefore, a key issue in strongly correlated electron systems 
is to understand if a ferromagnetic 
ground-state can survive at finite hole densities, finite Coulomb repulsions, or
can be stabilized by slight modifications to the pure Hubbard model. 

Recently there is a renewed interest in this field \cite{vollhardt} and many authors 
have focused on the two-dimensional infinite-$U$ Hubbard model 
\cite{doucot,kotliar,muller,edwards,putikka}, 
but the situation is still controversial.
In particular, Putikka {\it et al.} \cite{putikka}, using the 
high temperature expansion, found that the ground-state does not have a saturated 
magnetization for all hole densities. Due to the
uncertainties of their extrapolation at zero temperature, their
calculation cannot be considered a definite proof of the 
complete absence of a Nagaoka ground-state.
On the other hand, on any finite lattice with $L$ sites, the total spin $S$ of the 
ground-state is strongly dependent upon the number of holes and the boundary 
conditions chosen. For instance for two holes on any finite cluster with periodic
boundary conditions it is possible to show \cite{doucot} that the ground-state 
spin is not maximum, whereas for certain finite number of holes 
such that the ferromagnetic ground-state is nondegenerate with periodic boundary conditions,
the ground-state has maximum spin \cite{riera,barbieri}.
In general, if the boundary conditions are such that 
the ferromagnetic (paramagnetic)  state is a nondegenerate closed shell, 
ferromagnetism (paramagnetism) is energetically preferred.
Given the above difficulties, we believe that 
a reasonable approach to study the possible ferromagnetic instability in the infinite-$U$ 
Hubbard model is to study a finite density of holes and periodic boundary conditions such that 
the singlet state is nondegenerate, thus favoring the singlet state.
Whenever, for large size systems, we find that a state with a finite spin polarization has 
a lower energy than the singlet one, despite the fact that in the former case 
the closed shell condition may be not satisfied, this clearly implies a 
ferromagnetic instability weakly dependent on boundary conditions.
In order to proceed with this scheme, large cluster sizes are necessary since in small 
systems there are few and irrelevant singlet closed shell dopings.
In the present work we are able to consider fairly large lattice sizes 
so that our findings rely mainly on closed shell dopings, even though some of 
the calculations were performed with open shells, in order to interpolate 
between two closed shell densities. 

The Quantum Monte Carlo method (QMC) has recently been a subject of intense activity
\cite{nandini,runge,sorella1} due to its ability to handle the exponentially 
large Hilbert space dimension of strongly correlated systems and it is 
one of the few reasonable approaches for large size systems.
In this letter we have made extensive use of a recent QMC technique \cite{sorella2}, which is 
particularly suitable to estimating the ground-state energy of a Hamiltonian $H$ by applying a 
few Lanczos steps to a given starting variational wave function $|\Psi_G\rangle$.
By using the Green-function Monte Carlo with stochastic reconfiguration 
it is possible to compute the best energy state 
$|\Psi_p\rangle=\sum_{k=0}^p \alpha_k H^k |\Psi_G\rangle$, 
which defines the wave function with $p$ Lanczos steps applied to $|\Psi_G\rangle$.
This can be done, even for large size, at least up to $p=2$. 
Expectation values of the Hamiltonian $\langle \Psi_p | H | \Psi_p\rangle$ and corresponding 
energy variance $\sigma^2_p=\left[\langle \Psi_p | H^2 | \Psi_p\rangle 
-\langle \Psi_p | H | \Psi_p\rangle^2 \right]/L^2$, can then be computed by QMC, 
with high statistical accuracy.
As well known, an accurate estimate of the energy per site can be obtained by computing 
the energy as a function of the variance for different wave functions \cite{ceperley}. 
The variance is zero for an exact eigenstate and therefore, if different variational 
wave functions are systematically approaching the ground-state, the exact value of the energy
can be estimated by extrapolating the different calculations up to the zero variance limit.
The systematic approach of the Lanczos technique to the exact solution is 
remarkably accurate even for large system sizes. Thus the variance extrapolation 
technique combined with the Lanczos algorithm represents, at the moment, one of the most simple 
and efficient schemes to estimate the exact energy or, at least, the error of our best 
variational energy.
As shown in Ref.~\cite{sorella2}, with a reasonably good variational state,
it is possible to estimate exact energies within the statistical 
error, even for $L \sim 100$.

\begin{figure}
\centerline{\psfig{bbllx=50pt,bblly=210pt,bburx=530pt,bbury=670pt,%
figure=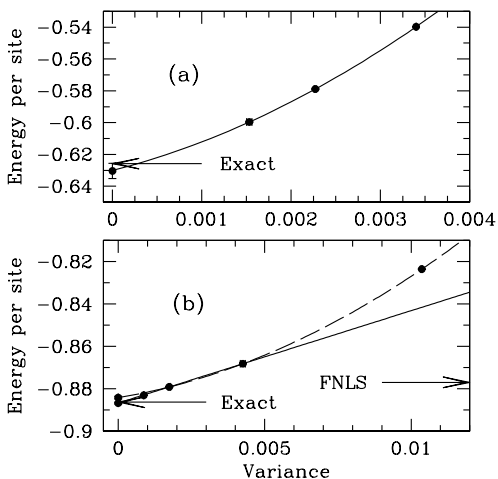,width=70mm,angle=0}}
\caption{\baselineskip .185in \label{check}
(a): variational energies for $p=0,1,2$ in the singlet subspace for 28
electrons on a one-dimensional chain of 50 sites. The continuous line is a quadratic fit.
The extrapolated point with zero variance is also shown.
(b): variational energies for $p=0,1,2,3$ in the singlet subspace for 10 electrons on 
a two-dimensional cluster of 18 sites. 
The continuous (long-dashed) line is a linear (quadratic) fit of the $p=1,2$ 
($p=0,1,2$) results. The extrapolated points with zero variance and the 
FNLS results are also shown.}
\end{figure}

For the present study, the Lanczos method also has the remarkable advantage that,
if $|\Psi_G\rangle$ has definite quantum numbers, the Lanczos method will conserve them exactly, 
and therefore it is possible to compute the lowest energy in the different spin sectors.
In order to show the quality of our best variational wave function on large size systems,
we will also use the standard fixed-node approximation \cite{fn}, by using the $p=1$ 
variational state as the guiding wave function, which will be denoted by FNLS. 
This approximation typically
leads to slightly lower energies than the best $p=2$ Lanczos ones.
However, within the 
fixed-node approach, the total spin is not conserved and therefore this approximation is not 
useful for estimating the spin dependence of the energy, which is our main task.
Moreover, within the fixed-node approach, it is not possible to calculate the energy variance and
therefore to assess the accuracy of the calculation.

\begin{figure}
\centerline{\psfig{bbllx=40pt,bblly=220pt,bburx=510pt,bbury=690pt,%
figure=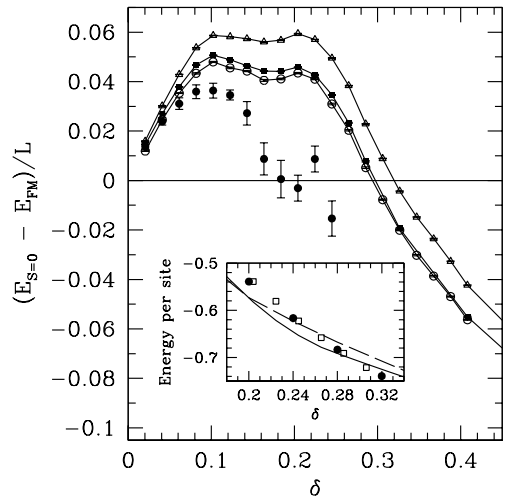,width=70mm,angle=0}}
\caption{\baselineskip .185in \label{98sites}
Differences between the singlet energy $E_{S=0}/L$ and the fully polarized one
$E_{FM}/L$ for the 98-site lattice as a function of the hole density for 
different approximations: $p=1$ (open triangles) and $p=2$ (full squares). 
The results for the FNLS approximation in the $S_z=0$ subspace 
(open circles) and the extrapolated values (full circles) are also reported.
The lines are guides to the eye.
In the inset: $E_{S=0}/L$ for the 50-site lattice (full circles) and for the 
98-site lattice (empty squares) are reported. The ferromagnetic energy for the 50-site
lattice (dashed line) and for the 98-site lattice (continuous line) are also shown.} 
\end{figure}

The $p=0$ variational wave function is
\begin{equation}\label{wf}
|\Psi_{G} \rangle = {\cal P}_N {\cal P}_G {\cal J} \prod_k
\left ( 1 + f_k c^{\dag}_{k,\uparrow} c^{\dag}_{-k,\downarrow} \right )
|0 \rangle,
\end{equation}
where ${\cal P}_N$ is the projector onto the subspace of $N$ particles,
${\cal P}_G$ is the Gutzwiller projector, which forbids doubly
occupied sites, ${\cal J} ={\rm exp}(\gamma/2 \sum_{i,j} v_{i,j} h_i h_j )$
is a Jastrow factor,
defined in term of the hole density at site $i$ $h_i= (1-n_{i\uparrow}) (1-n_{i,\downarrow})$,
and $\gamma$ is a variational parameter.
For the potential $v$, we take the exact analytic form 
determined by considering the holes as hard-core bosons at the same density
\cite{francjic}.
The variational parameters $f_k$ of the pairing wave function
may also represent the Gutzwiller wave function in the 
particular case of $f_k= \Theta (\epsilon_F - \epsilon_k)$,
where $\epsilon_F$ is the free Fermi energy, $\epsilon_k$ is the corresponding 
dispersion $\epsilon_k=-2 t (\cos k_x  + \cos k_y)$ and $\Theta$ is the step function.
For closed shell fillings, this wave function is a singlet.
For open shell,
we take $f_k$ with a small d-wave or s-wave symmetry in order to split the 
bare degeneracy, the resulting wave function being a singlet. Analogously,
by using the particle-hole transformation on down spins,
we are able to consider a number of up electrons which is different from the number of 
down electrons, corresponding to a finite total spin $S>0$.
We use the $45^\circ$ tilted squares with $L=l\sqrt{2}\times l\sqrt{2}$ sites
with periodic boundary conditions, in order to have the full spatial 
symmetries of the infinite lattice and remove the huge degeneracy of the conventional 
$l \times l$ square lattices near half-filling.

For the infinite-$U$ Hubbard model, we have found that the best variational wave function
of the class given in Eq.~(\ref{wf}) has a free electron determinant, 
$f_k=\Theta(\epsilon_F - \epsilon_k)$, even for nonzero total spin.
The Jastrow term ${\cal J}$ provides a sizable
improvement in the variational energy (of order $0.01$ per site). 
Definitely, this class of wave functions is particularly accurate in the large doping limit 
where the paramagnetic state is expected to be stable apart from 
Kohn-Luttinger superconducting instabilities, which are however expected 
to lead to negligible energy gain and are irrelevant for our purposes.  
In the more delicate small-doping region, based on variational QMC
calculations \cite{shiba}, it appears that if 
nontrivial order parameters, other than the ferromagnetic one, 
show up in the ground-state, they should lead 
to a marginal energy gain. Within this assumption, even in this case, 
the Gutzwiller wave function can be considered a good starting point 
for our projection technique. 
 
In order to test the Lanczos step technique and apply it to the infinite-$U$ Hubbard model, 
we show in Fig.~\ref{check} the energy as a function of the variance 
for two closed shell cases where the exact values are known: a small two-dimensional 
18-site lattice, and a larger one-dimensional chain.
The approach to the exact ground-state energy is remarkably good, even when, 
as shown in Fig.~\ref{check}(a), the $p=0$ variational wave function is about 
$20\%$ off in energy in the one-dimensional case, where the exact solution is known
by Bethe ansatz. For small two-dimensional systems, 
the $p=0$ variational wave function is zero for several electronic 
configurations of the Hilbert space, whereas, as soon as we apply the first Lanczos step, 
this pathology is substantially removed.
Thus, in this case, it is natural to fit linearly only the $p=1,2$ results,
the extraplated value being not affected by the inclusion of the $p=3$ point
(which is possible for this small cluster), see Fig.~\ref{check}(b).
Nevertheless, even by fitting quadratically the $p=0,1,2$ points, we obtain a
rather accurate result.
On larger size systems, $L \gtrsim 50$, we have verified that the number of 
zeros in $|\Psi_G\rangle$ is 
negligible and that all the three energies can be safely included in the same fit.

We perform a systematic study of the spin instability and we
consider the 50-site and the 98-site lattices. 
The results for these two systems are both in qualitative and quantitative
agreement, suggesting that, for $L \sim 100$, the finite size corrections are 
small compared to the energy difference between the singlet and the saturated ferromagnet.
In Fig.~\ref{98sites} we report the
difference in energy between the singlet and the fully polarized
state for different hole densities and QMC techniques for the 98 sites.
For large doping the homogeneous singlet state is energetically well below
the ferromagnet but, by decreasing the number of holes, there is
a clear evidence of a fully polarized ground-state.
Indeed all the Monte Carlo methods (variational with $p=0,1,2$ and FNLS)
predict that the singlet energy is below the fully polarized one only for $\delta > \delta_{SF}$,  
where the critical doping $\delta_{SF} \sim 0.28 \div 0.30$. 
Of course, $\delta_{SF}$ gives an upper bound for the critical density $\delta_{c_1}$, 
below which the saturated ferromagnetism is stable. 
Indeed, unsaturated ferromagnetism can appear below $\delta_{SF}$. 

\begin{figure}
\centerline{\psfig{bbllx=50pt,bblly=220pt,bburx=530pt,bbury=640pt,%
figure=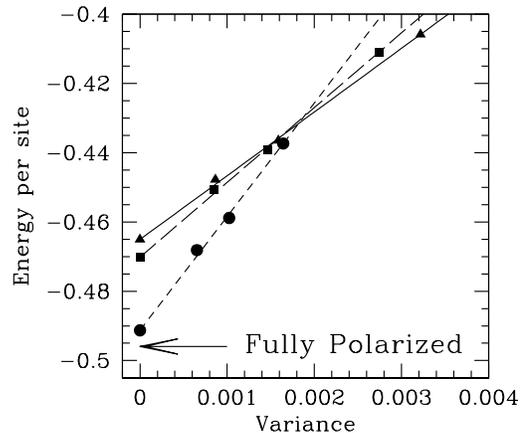,width=70mm,angle=0}}
\caption{\baselineskip .185in \label{50sites}
Variational energies with $p=0,1,2$ in different total spin sectors for 42 electrons
on 50 sites: $S=0$ (triangles, continuous line), $S=8$ (squares, long-dashed 
line), and $S=16$ (circles, short-dashed line).
The fully polarized energy for the same lattice is also reported.} 
\end{figure}

If we consider the extrapolated value of the energy, $\delta_{SF}$ is
pushed at much lower doping $\delta^{ext}_{SF} \sim 0.20$. This transition point 
appears to be in very good agreement with the best variational estimate for the 
instability of the fully polarized state \cite{muller,edwards}. 
In our numerical calculation we are able to study directly the instability of the 
singlet-paramagnetic state, starting from the ``safe'' large doping limit.  
The fact that we have obtained a value for the transition point
very similar to the one found in Refs.~\cite{muller,edwards}, supports the 
validity of both approaches in predicting a stable ferromagnetic region.

In order to show that the fully polarized state has the lowest energy in the 
low-doping regime, in Fig.~\ref{50sites} we show the variational 
energies for 42 electrons on 50 sites, corresponding to $\delta=0.16 < \delta_{SF}$. 
We consider all the possible values of the spin which fulfill the closed shell condition,
whereas the fully polarized state is not a closed shell.
This condition, as discussed in the introduction, certainly frustrates the fully polarized 
state and favors the singlet. Moreover, we have considered a linear fit of the $p=0,1,2$
energy results, yielding a slightly lower extrapolated energy compared to the one 
obtained with the more accurate quadratic fit. 
Even with this choice, the ferromagnetic state remains clearly stable.
Moreover, the energy decreases monotonically with increasing total spin.
The outcome of this calculation is that, as we increase the system size, 
the ferromagnetic state becomes less and less frustrated.
For the singlet state instead, the boundary conditions do not play a crucial 
role and the corresponding energy per site is very weakly size dependent 
(see inset of Fig.~\ref{98sites}).

\begin{figure}
\centerline{\psfig{bbllx=50pt,bblly=220pt,bburx=530pt,bbury=670pt,%
figure=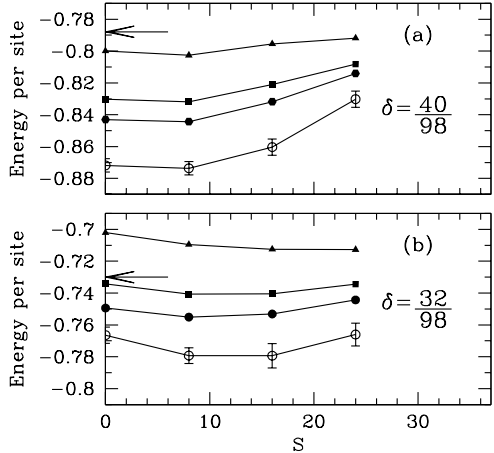,width=70mm,angle=0}}
\caption{\baselineskip .185in \label{minimum}
(a): variational results with $p=0$ (full triangles), $p=1$ (full squares) and
$p=2$ (full circles) for the ground-state energy as a 
function of the total spin $S$ for 58 electrons on 98 sites. The extrapolated energies
(empty circles) are also shown. The ferromagnetic energy
is marked by the arrow and the lines are guides to the eye. 
(b): the same for 66 electrons on 98 sites.}
\end{figure}

Since in the low-doping region the fully polarized state is found
to be stable, it is reasonable to expect a partially polarized 
ground-state for intermediate doping.
For very large doping, $\delta > 0.50$, the ground-state is found to be
paramagnetic and the energy is monotonically increasing with increasing
total spin, implying a finite spin susceptibility.
For every variational Monte Carlo calculation the results are
consistent with a paramagnetic ground-state and a finite spin susceptibility.
By decreasing the hole density, the low-spin states collapse to the singlet ground-state
and eventually become degenerate with it. For 40 holes on 98 sites, we found that,
for all the Monte Carlo techniques used, the singlet is degenerate, within the 
statistical error, with the $S=8$ state, both the states being closed shell,
see Fig.~\ref{minimum}(a). 
Indeed, although in the variational calculation there
is a small but significant difference between the energy of the $S=0$ and the 
$S=8$ state, by increasing the accuracy, with $p$ Lanczos steps, the two energies tend
to become closer and closer and even the extrapolated values give a degeneracy of these two 
spin sectors. This indicates that at doping $\delta_{c_2} \sim  0.40$ the uniform 
spin susceptibility diverges and a finite magnetic moment sets in for $\delta < \delta_{c_2}$.
These results strongly support a second-order transition to a partially polarized ferromagnet, 
and a nonzero magnetic moment is in fact seen at doping $\delta \sim 0.33$, 
where the energy is minimum for $S \sim 10$, see Fig.~\ref{minimum}(b).

To conclude, unsaturated ferromagnetism begins already at $\delta \sim 0.40$ and
rather reliable numerical evidence is given that Nagaoka ferromagnetism is 
not simply an irrelevant property of a single hole, but is realistic physical behavior of 
strongly correlated electrons.
Since at small but finite doping ferromagnetism is stabilized by a macroscopic 
energy gain, this property is robust against small perturbation of the model,
such as a large repulsion $U$ or a small positive coupling $J$ in the $t-J$ model
\cite{note2}. 

We acknowledge M. Capone, L. Capriotti, C. Castellani, A. Parola for useful discussions, 
and P. Tangney for a careful reading of the manuscript. 
This work was partially supported by MURST (COFIN99).


\begin{references}
\bibitem{note2} Preliminary results confirm that for $\delta \sim 0.1$ and
   $J/t \lesssim 0.05$, the ground-state is ferromagnetic, whereas, by decreasing $\delta$,
   first ferromagnetism melts, and then, at very small doping, antiferromagnetism
   is finally stabilized.
\end{references}
\end{document}